\documentclass[12pt]{article}
\usepackage{amsmath}
\usepackage{amssymb}
\usepackage{cite}

\makeatletter

\begin{document}
\begin{titlepage}
\thispagestyle{empty}

\bigskip

\begin{center}
\noindent{\Large \textbf
{RNS and Pure Spinors Equivalence for Type I Tree Level Amplitudes Involving up to Four Fermions}}\\

\vspace{0,5cm}

\noindent{G. Alencar ${}^{a}$\footnote{e-mail: geovamaciel@gmail.com }, M. O. Tahim ${}^{a}$, R. R. Landim ${}^{b}$ and R.N. Costa Filho ${}^{b}$}

\vspace{0,5cm}

{\it ${}^a$Universidade Estadual do Cear\'a, Faculdade de Educa\c c\~ao, Ci\^encias e Letras do Sert\~ao Central- 
R. Epit�cio Pessoa, 2554, 63.900-000  Quixad\'{a}, Cear\'{a},  Brazil.}

\vspace{0.2cm}

 {\it ${}^b$Departamento de F\'{\i}sica, Universidade Federal do Cear\'{a}-
Caixa Postal 6030, Campus do Pici, 60455-760, Fortaleza, Cear\'{a}, Brazil. 
 }

\end{center}

\vspace{0.3cm}

\begin{abstract}
  
In this paper we give a proof of the equivalence between the RNS and Pure Spinor formalism for Type I tree level amplitudes involving up to four fermions. This result have been obtained
previously for amplitudes involving only closed or open amplitudes and here we extend it to include Type I amplitudes. For this we first prove the cyclic symmetry of Type I superstring in a way that
it is also valid for the Pure Spinor formalism. The technique used is applied to simplify the 
calculation of the tree level three point amplitude previously computed. As a byproduct, we are able to calculate the gauge anomaly 
amplitude of type I Superstrings in this formalism.
\end{abstract}


\end{titlepage}

\section{Introduction}

The covariant quantization of superstring theory have been an unsolved
problem for a long time. The covariant quantization presents manifest
supersymmetry and gives us an explicit covariant prescription to compute
tree-level and higher loop scattering amplitudes for an arbitrary
number of states \cite{Nathan 2000}. This is important for understanding the low energy
limit of superstrings through the construction of effective actions
corresponding to such amplitudes. Besides, it is possible to make
explicit calculations in a specific background, a powerful characteristic
that is applied to understand the $AdS/CFT$ correspondence from the
string theory side \cite{Berkovits:2004xu,Vallilo:2003nx,Vallilo:2002mh,Bianchi:2006im,Berkovits:2007zk,Berkovits:2000yr,Bedoya:2010qz}.
Beyond this background, the formalism have been used to study sigma
models in \cite{Bedoya:2006ic,Bedoya:2008mz,Bedoya:2008yw}. This
new formalism keeps all the good properties of Ramond-Neveu-Schwarz
and Green-Schwarz and does not have its undesired characteristics.
In the Ramond-Neveu-Schwarz formalism, when the number of loops in
computations of scattering amplitudes are increased, more and more
spin structures have to be considered making the computations
very long. On the other hand, in the Green-Schwarz formalism the quantization
is only possible in the light-cone gauge and the amplitude computations
involve non-covariant operators at the interaction points.

In order to establish the power of this method it is important to
check its consistency by simply comparing its results to those coming
from the standard Ramond-Neveu-Schwarz (RNS) an Green-Schwarz (GS)
formulations. In this sense, several important results related to
scattering amplitude computations were achieved. The tree-level amplitudes
were shown to be equivalent with the RNS computations in \cite{Berkovits:2000ph},
for amplitudes containing any number of bosons and up to four fermions.
Years later, the multiloop prescription was given \cite{Berkovits:2004px,Berkovits:2005df}
showing the equivalence up to two-loop level. The explicit calculations
of equivalence of results for one- and two-loop amplitudes with the
minimal and non-minimal formalisms, the computation of the gauge variation
of the massless six-point open string amplitude, obtaining the kinematic
factor related to the anomaly cancellation, between other important
related results were established \cite{mafra1loop,mafra2loops,mafraidentities,Berkovits:2006bk,Mafra:2009bz,Mafra:2010ir}.
More recently, the tree level amplitude of six massless open strings
was obtained \cite{Mafra:2010gj}. A recursive formula for super Yang-Mills
color-ordered $n$ - point tree amplitudes based on the cohomology
of pure spinor superspace in ten space-time dimensions was constructed
\cite{Mafra:2010jq}. Another example of the power of the formalism
is the computation of the coefficient of the massless one-loop and
two-loop four-point amplitude from first principles \cite{Gomez:2009qd,Gomez:2010ad},
which was not possible with RNS and GS. 

The interesting thing here is that type I supergravity can be described
in quantum language with the use of superspace, therefore solving
two important problems. We know, for example, that all the cubic terms are encoded in an unique and simple superspace expression \cite{Alencar:2008fy}, showing the great simplification brought by the formalism.
Hence, many terms of the effective action are obtained when we take this expression and expand in components. From the standpoint of field theory, the effective action for the type I supergravity is obtained from 
the global super Yang-Mills action
by imposing local supersymmetry. This procedure generates many compensation
terms \cite{Bergshoeff:1981um} that are interpreted as interaction terms. From the standpoint of superstrings, all these interaction terms must come out naturally from amplitude computations. 
The interaction terms of the effective action for type
I supergravity have some interesting properties. A very peculiar one is the fact that there is a coupling between the Kalb-Ramond and two photons. This term is necessary inorder to garantee local supersymmetry. 
In order to keep gauge invariance, the Kalb-Ramond field must have a unusual transformation under U(1) symmetry. This coupling will become very
important for the mixed anomaly cancellation in the SO(32) theory.

In all of these calculations we strongly use the cyclic symmetry property of scattering amplitudes. This means basically that the tree amplitudes do not depend on which
of the vertex operators are chosen to be unintegrated. In a
more specific result \cite{Alencar:2008fy} this symmetry was used to compute the effective action
for type I supergravity using the pure spinor formalism
for superstrings. That work regarded the tree level approximation
for Type I superstring amplitudes. If we identify the fermionic and bosonic vertex operators and use the cyclic symmetry, then that computation can be simplified.
As we will see, the bosonic vertex operators contributes to
zero or one thetas and therefore will always increase the number of
thetas. However the expression for the fermionic integrated vertex operator acts as a theta derivative,
and therefore contributes to $-1$ in the computation of the five
thetas,  afact which can make the computations very long. On the other hand
the unintegrated contributes with two thetas. Therefore the strategy
is to use the cyclic symmetry to always choose the fermion operators
as unintegrated ones. 

In this paper, after an explanation about the cyclic symmetry for Type I tree level amplitudes, we are interested in showing the equivalence between the RNS and Pure Spinor formalism for Type I amplitudes involving up to four fermions.
We also show how the techniques used here can simplify the computation of the three point tree level amplitude.
As a byproduct, the computation of the gauge anomaly amplitude of type I Superstrings from the viewpoint of pure spinors is obtained.
The organization of the work is as follows: in the second section we give a review of the scattering amplitude prescription in the pure spinor formalism. In the third section we give the proof of cyclic symmetry. 
In the fourth, fifth and sixth sections we do explicit computations of amplitude equivalences. In the seventh section we simplify the computation for the three point type I amplitude, discussing the 
appearance of the well known gauge anomaly in this model. In the final section we present conclusions and perspectives.

\section{The Pure Spinor Formalism}

\subsection{Tree-level Amplitudes in Type I Pure Spinor Formalism}

The pure spinor formalism \cite{Nathan2000} of superstrings contains
the usual bosonic $X$ field, the spinor field $\theta$ and a pure
spinor $\lambda$ and its respective conjugate momenta $p,\omega$.
The Lorentz generators for the ghosts $\lambda$ are

\[
N^{nm}=\frac{\alpha^{\prime}\left(\lambda\gamma^{mn}\omega\right)}{4}.
\]
and the OPEs are given by

\[
p_{\alpha}(z_{1})\theta^{\beta}(z_{2})=-\frac{\delta_{\alpha}^{\beta}}{z_{2}-z_{1}},\;\bar{p}_{\alpha}(\bar{z}_{1})\bar{\theta}^{\beta}(\bar{z}_{2})=-\frac{\delta_{\alpha}^{\beta}}{\bar{z}_{2}-\bar{z}_{1}}.
\]
Here we must be careful because we will consider Type I amplitudes with
Riemann surface given by the upper half complex plane. Therefore we also have
the following OPEs 

\[
p_{\alpha}(z_{1})\bar{\theta}^{\beta}(\bar{z}_{2})=-\frac{\delta_{\alpha}^{\beta}}{\bar{z}_{2}-z_{1}},\;\bar{p}_{\alpha}(\bar{z}_{1})\theta^{\beta}(z_{2})=-\frac{\delta_{\alpha}^{\beta}}{z_{2}-\bar{z}_{1}},
\]
due to the mixing between the left and right movers. For the $X_{\mu}$
field we have

\begin{equation}
:X^{\mu}\left(z_{1}\right)X^{\nu}\left(z_{2}\right):\sim-\frac{\alpha^{\prime}}{2}\eta^{\mu\nu}\left[\ln\left\vert z_{1}-z_{2}\right\vert ^{2}+\ln\left\vert z_{1}-\bar{z}_{2}\right\vert ^{2}\right].
\end{equation}
This formalism is symmetric under a BRST transformation for the left
and right moving sectors, with generators given by

\begin{equation}
Q_{c}=\int dz\lambda^{\alpha}d_{\alpha}^{c},\quad\bar{Q}_{c}=\int dz\bar{\lambda}^{\alpha}\bar{d}_{\alpha}^{c}\label{operador BRST},
\end{equation}
with the definition

\[
d_{\alpha}^{c}=\frac{\alpha^{\prime}}{2}p_{\alpha}-\theta\gamma^{m}\partial x_{m}-\frac{1}{8}\gamma_{\alpha\beta}^{m}\gamma_{m\delta\eta}\theta^{\beta}\theta^{\delta}\partial\theta^{\eta},
\]
and the similarly for $\bar{d}_{\alpha}^{c}$. With this we
can compute the following important OPEs

\begin{equation}
d_{\alpha}^{c}\left(z_{i}\right)V\left(z_{j}\right)\sim-\frac{\alpha^{\prime}}{2}\frac{D_{\alpha}V}{z_{j}-z_{i}}-\frac{\alpha^{\prime}}{2}\frac{\bar{D}_{\alpha}^{\prime}V}{\bar{z}_{j}-z_{i}}\label{ope d v},
\end{equation}
and

\begin{equation}
\bar{d}_{\alpha}^{c}\left(z_{i}\right)V\left(z_{j}\right)\sim-\frac{\alpha^{\prime}}{2}\frac{D_{\alpha}^{\prime}V}{z_{j}-z_{i}}-\frac{\alpha^{\prime}}{2}\frac{\bar{D}_{\alpha}V}{\bar{z}_{j}-z_{i}}\label{ope bar{d} v-1},
\end{equation}

\[
d_{\alpha}\left(z_{i}\right)d_{\beta}\left(z_{j}\right)\sim\frac{\alpha^{\prime}}{2}\frac{\gamma_{\alpha\beta}^{m}\Pi_{m}}{z_{j}-z_{i}}.
\]
where

\[
D_{\alpha}=\frac{\alpha^{\prime}}{2}\partial_{\alpha}+\theta\gamma^{m}\partial_{m},\ \ \bar{D}_{\alpha}=\frac{\alpha^{\prime}}{2}\bar{\partial}_{\alpha}+\bar{\theta}\gamma^{m}\partial_{m}.
\]
From the last OPE we see that

\[
Q_{c}^{2}=-\frac{\alpha^{\prime}}{2}\oint\lambda^{\alpha}\lambda^{\beta}\gamma_{\alpha\beta}^{m}\Pi_{m},\;\bar{Q}_{c}^{2}=-\frac{\alpha^{\prime}}{2}\oint\bar{\lambda}^{\alpha}\bar{\lambda}^{\beta}\gamma_{\alpha\beta}^{m}\bar{\Pi}_{m},\; Q_{c}\bar{Q}_{c}=0,
\]
and therefore the BRST operators are nilpotent only if they satisfy
the condition

\begin{equation}
\lambda\gamma^{m}\lambda=\bar{\lambda}\gamma^{m}\bar{\lambda}=0.\label{vinculo}
\end{equation}

A spinor that satisfies the above condition was called a pure spinor
by Cartan \cite{cartan}. The only OPE involving ghost fields which will be needed
in this work is\cite{Nathan2000}

\begin{equation}
N^{mn}\left(z_{1}\right)\lambda^{\alpha}\left(z_{2}\right)=\frac{\alpha^{\prime}}{4\left(z_{2}-z_{1}\right)}\left(\lambda\gamma^{mn}\right)^{\alpha}.\label{OPE N lambda}
\end{equation}
With the BRST operator we can construct the vertex operator.
The fixed one is defined as the cohomology with ghost number $+1$.
The most simple object of ghost number one is given by

\[
V=g_{o}^{\prime}\lambda^{\alpha}A_{\alpha}(z,\theta),
\]
where $A_{\alpha}$ is a spinorial superfield. The physical state
condition

\[
QV=0,
\]
 give us the equations of motion

\begin{equation}
D_{\alpha}A_{\beta}+D_{\beta}A_{\alpha}=\gamma_{\alpha\beta}^{m}A_{m},\label{Eq of motion}
\end{equation}
and $A_{m}$ is a vector superfield. These are the right equations
of supergravity. A fact that will be used in this work is the definition
of the integrated operators. For the open string case this is defined
as $[Q_o,U]=\dot{V}$ and for the closed $\{Q_c,[\bar{Q_c},U]\}=\partial\bar{\partial}V$. The important fact here is that, for Type I strings case,  we have the
definition
\begin{equation}
Q_oU(z,\bar{z})=Q_o(e^{ikX}U(\theta)\bar{U}(\bar{\theta}))=\{\partial(e^{ikX}V(\theta))\bar{U}(\bar{\theta})+U(\theta)\bar{\partial}(e^{ikX}\bar{V}(\bar{\theta}))\}\label{fixo}
\end{equation}
what give us a `half integrated" operator.
In the above expression $U(\theta)\bar{U}(\bar{\theta})$ are functions
that depend only on $\theta$ and $\bar{\theta}$ respectively.

The integrated vertex operator for the
open string, in its turn, has ghost number zero and is 
\[
g_{o}^{\prime}\int dy_{3}\left(\partial\theta^{\alpha}A_{\alpha}+A_{m}\Pi^{m}+d_{\alpha}W^{\alpha}+\frac{1}{2}N^{nm}\mathcal{F}_{nm}\right),
\]
where $N^{mn}$ was defined before and $A_{\alpha}$, $A_{m}$ and
$d_{\alpha}$ are defined above. $W^{\alpha}$ and $F_{mn}$ are
field strengths given by

\[
W^{\alpha}=\frac{1}{10}\gamma_{m}^{\alpha\beta}D_{\beta}A^{m},\ \ \mathcal{F}_{mn}=2\partial_{\lbrack m}A_{n]}.
\]
When necessary, the superfields will be expanded in components. The
vertex operator for the closed string is given by the product of two
open string operators $\ensuremath{\lambda^{\alpha}A_{\alpha}}\ensuremath{\bar{\lambda}^{\alpha}\bar{A}_{\alpha}}$.
The prescription to compute tree-level closed string amplitudes with
the pure spinor formalism is given by 
\begin{equation}
\mathcal{A}_{n}=\langle V^{1}(z_{1})V^{2}(z_{2})V^{3}(z_{3})\int d^{2}z_{4}U_{4}(z_{4})...\int d^{2}z_{N}U_{N}(z_{N})\rangle
\end{equation}
and after the necessary $\theta$-expansions of superfields and use
of OPEs, the integration of the zero-modes of $\lambda^{\alpha}$
and $\theta^{\alpha}$ is carried out by taking only the terms which
contain three $\lambda$s and five $\theta$s in the correlator which
are proportional to the pure spinor measure 
\[
\left\langle \left(\lambda\gamma^{m}\theta\right)\left(\lambda\gamma^{n}\theta\right)\left(\lambda\gamma^{p}\theta\right)\left(\theta\gamma_{mnp}\theta\right)\right\rangle =c.
\]

Here the normalization can be chosen in order to obtain the right
results for comparison with other formalisms. We also have
use two different values for $c$ in order to compare our result with
the RNS formalism and with a previous computation of the three point
function in the pure spinor formalism.

\subsection{Bosonic and Fermionic Vertex Operators }

As said in the introduction, we must identify the bosonic and fermionic
vertex operators. These have been found in \cite{Berkovits:2000ph} and we 
need to be careful with the factors of $\alpha^{\prime}$. These operators
are given by 
\[
U^{B}=a^{m}[\partial x_{m}-ik^{n}(N_{mn}-\frac{\alpha^{\prime}}{4}p\gamma_{mn}\theta)+2\theta`s...]e^{ik\cdot x},\quad U^{F}=-\xi^{\alpha}[\frac{\alpha^{\prime}}{2}p_{\alpha}+...]e^{ik\cdot x},
\]
 
\[
V^{B}=a^{m}[\frac{1}{2}\lambda\gamma_{m}\theta+3\theta`s...]e^{ik\cdot x},\quad V^{F}=\xi^{\alpha}[\frac{1}{3}(\lambda\gamma_{m}\theta)(\gamma^{m}\theta)_{\alpha}+4\theta`s...]e^{ik\cdot x},
\]
where in the above expressions, $U,V$ stands for integrated and unintegrated
operators respectively. From the above expression we can see that
the fermionic integrated vertex operator acts as a theta derivative,
and therefore contributes to $-1$ in the computation of the five
thetas, what can make the computations very long. On the other hand
the unintegrated contributes with two thetas. Therefore the strategy
here is to use the cyclic symmetry to always choose the fermion operators
as unintegrated ones. The bosonic vertex operators contributes to
zero or one thetas and therefore will always increase the number of
thetas. With this strategy we will see that only the above order in
thetas are needed. As a simplification we must use the definitions
\[
b_{m}\equiv\lambda\gamma_{m}\theta,M_{mn}\equiv N_{mn}-\frac{\alpha^{\prime}}{4}(p\gamma_{mn}\theta)
\]
and 
\[
f_{\alpha}\equiv(\lambda\gamma^{m}\theta)(\gamma_{m}\theta)_{\alpha}.
\]

With this we have 
\[
U^{B}=a^{m}[\partial x_{m}-ik^{n}M_{mn}+2\theta`s...]e^{ik\cdot x},\quad U^{F}=-\xi^{\alpha}[\frac{\alpha^{\prime}}{2}p_{\alpha}+...]e^{ik\cdot x},
\]
 
\[
V^{B}=a^{m}[\frac{1}{2}b_{m}+3\theta`s...]e^{ik\cdot x},\quad V^{F}=\xi^{\alpha}[\frac{1}{3}f_{\alpha}+4\theta`s...]e^{ik\cdot x}.
\]
and using our previous OPEs we get

\[
M_{mn}(z_{1})f_{\alpha}(z_{2})\to\frac{\alpha^{\prime}(\gamma_{mn})_{\alpha}{}^{\beta}f_{\beta}(z_{2})}{4(z_{1}-z_{2})},\quad M_{mn}(z_{1})b_{p}(z_{2})\to\alpha^{\prime}\frac{\eta_{np}b_{m}(z_{2})-\eta_{mp}b_{n}(z_{2})}{2(z_{1}-z_{2})}.
\]

The other OPEs needed in this work are

\begin{equation}
M_{kl}(z_{1})M_{mn}(z_{2})=\frac{\alpha^{\prime}}{2}\frac{\eta_{m[l}M_{k]n}(z_{2})-\eta_{n[l}M_{k]m}(z_{2})}{z_{1}-z_{2}}+\frac{\alpha^{\prime2}}{4}\frac{\eta_{kn}\eta_{lm}-\eta_{km}\eta_{ln}}{(z_{1}-z_{2})^{2}}
\end{equation}
and

\begin{equation}
M_{mn}(z_{1})p_{\alpha}(z_{2})\to\frac{\alpha^{\prime}(\gamma_{mn})_{\alpha}{}^{\beta}p_{\beta}(z_{2})}{4(z_{1}-z_{2})}.
\end{equation}

These OPEs will be needed for comparison with the respective ones in
the RNS formalism.

\section{Cyclic Symmetry for Type I Superstring Amplitudes}

In this section we present a proof of the superstring cyclic symmetry for tree level Type I amplitudes that is valid for the pure spinor formalism.
As said in the introduction, the Type I effective action for supergravity was recently computed in the
framework of pure spinor superstrings and the tree-level Type I amplitudes (for
one closed and two open strings) were calculated. In these calculations, the cyclic symmetry
was used as a tool but a proof was not presented. This exists within
the BRST formalism such that it also applies to pure spinors, but it
only works for the open string or closed string amplitudes\cite{Berkovits:2000ph}. We extend this to the general case including Type I strings. This symmetry will also be
needed to prove the RNS and Pure Spinor equivalence for amplitudes involving up to four fermions.
\subsection{Cyclic Symmetry for Open Strings}

For the sake of simplicity, we first give a proof for the cyclic symmetry for open strings
tree amplitudes, a result that is similar to the one given in \cite{Berkovits:2000ph}
for closed strings. The amplitude in this case is given by 
\begin{equation}
\langle V_{1}(y_{1})V_{2}(y_{2})V_{3}(y_{3})\int_{y_{3}}^{y_{1}}dy_{4}U_{4}(y_{4})\int_{y_{4}}^{y_{1}}dy_{5}U_{5}(y_{5})...\int_{y_{N-1}}^{y_{1}}d^{2}y_{N}U_{N}(y_{N})\rangle.
\end{equation}
Considering the delta function representation $\dot{\Theta}(y-y^{\prime})=\delta(y-y^{\prime})$,
where $\Theta$ is the Heaviside step function, and using $[Q,U(y)]=\dot{V}$
we obtain 
\begin{equation}
\int_{y_{2}}^{y_{3}}dy\{\Theta(y-y_{2})-\Theta(y-y_{3})+\Theta(y_{3}-y_{4})\}[Q,U_{3}(y)]=V_{3}(y_{3})-V_{3}(y_{2}).
\end{equation}
 Now using the cancelled propagator argument (CPA) \cite{polchinski},
the second term in the rhs can be disregarded and we get 
\begin{eqnarray}
\langle V_{1}(y_{1})V_{2}(y_{2})\int_{y_{2}}^{y_{3}}dy\int_{y_{3}}^{y_{1}}dy_{4}dy\{\Theta(y-y_{2})-\Theta(y-y_{3})+\Theta(y_{3}-y_{4})\}\times\nonumber \\
\times[Q,U_{3}(y)]U_{4}(y_{4})\int_{y_{4}}^{y_{1}}dy_{5}U_{5}(y_{5})...\int_{y_{N-1}}^{y_{1}}d^{2}y_{N}U_{N}(y_{N})\rangle
\end{eqnarray}
 for the amplitude.

Now we pull off the BRST operator to all the other vertex. Again the
CPA cancels the contribution of all of them and we are left only with
\begin{eqnarray}
\langle V_{1}(y_{1})V_{2}(y_{2})\int_{y_{2}}^{y_{3}}dyU_{3}(y)\int_{y_{3}}^{y_{1}}dy_{4}\{\Theta(y-y_{2})-\Theta(y-y_{3})+\Theta(y_{3}-y_{4})\}\times\nonumber \\
\times[Q,U_{4}(y_{4})]\int_{y_{4}}^{y_{1}}dy_{5}U_{5}(y_{5})...\int_{y_{N-1}}^{y_{1}}d^{2}y_{N}U_{N}(y_{N})\rangle,
\end{eqnarray}
leading to
\begin{equation}
\langle V_{1}(y_{1})V_{2}(y_{2})\int_{y_{2}}^{y_{3}}dyU_{3}(y)V_{4}(y_{3})\int_{y_{4}}^{y_{1}}dy_{5}U_{5}(y_{5})...\int_{y_{N-1}}^{y_{1}}d^{2}y_{N}U_{N}(y_{N})\rangle.
\end{equation}

The same argument applied here will be used to the case of Type I amplitudes.

\subsection{Cyclic Symmetry for Type I Strings}

Here we give the proof for the cyclic symmetry for Type I amplitudes. We
use the same sort of argument used previously for the open string
with the $\Theta$ function. We must point that this method is necessary
here because differently from \cite{Berkovits:2000ph} we can have a half fixed operator.
This does not happen in the non-mixing string case. Using
the steps above we write the amplitude involving three open
fixed operators as 
\begin{eqnarray}
\langle V_{1}(y_{1})V_{2}(y_{2})V_{3}(y_{3})\int dz_{4}d\bar{z}_{4}U_{4}(z_{4},\bar{z}_{4})\times\nonumber \\
\times\int d^{2}z_{5}U_{5}(z_{5},\bar{z}_{5})...\int d^{2}z_{N}U_{N}(z_{N},\bar{z}_{N})\rangle.
\end{eqnarray}
The dots can include any number of open or closed strings. In the above amplitude we consider only three open strings fixed and
all the closed strings integrated, because here we are concerned only with the exchange between open and closed strings. Using the same argument as before we get 
\begin{align}
\langle V_{1}(y_{1})V_{2}(y_{2}) & \int_{y_{2}}^{y_{3}}dy\int dz_{4}d\bar{z}_{4}U_{4}(z_{4})\{\Theta(y-y_{2})-\Theta(y-y_{3})+\ln(\left|z_{3}-z_{4}\right|^{2})\}\times\nonumber \\
 & [Q,U_{3}(y)]\bar{U}_{4}(\bar{z}_{4})\int d^{2}z_{5}U_{5}(z_{5},\bar{z}_{5})...\int d^{2}z_{N}U_{N}(z_{N},\bar{z}_{N})\rangle.
\end{align}
The function $\ln(\left|z_{3}-z_{4}\right|^{2})$ was chosen such that
we have a well defined real $y$ integration. Using BRST invariance
of the amplitude to pull it off to the other vertex operators and
using again the CPA, the only remaining term is the one coming from $U_{4}(z_{4},\bar{z}_{4})$.
Therefore because of (\ref{fixo}) we get 
\begin{align}
\langle V_{1}(y_{1})V_{2}(y_{2})\int_{y_{2}}^{y_{3}}dy\int dz_{4}d\bar{z}_{4}U_{3}(y)\{\Theta(y-y_{2})-\Theta(y-y_{3})+\ln(\left|z_{3}-z_{4}\right|^{2})\}\times\nonumber \\
\{\partial(e^{ikX}V(\theta))\bar{U}(\bar{\theta})+U(\theta)\bar{\partial}(e^{ikX}\bar{V}(\bar{\theta}))\}\int d^{2}z_{5}U_{5}(z_{5},\bar{z}_{5})...\int d^{2}z_{N}U_{N}(z_{N},\bar{z}_{N})\rangle.
\end{align}
 Performing the integration we finally get 
\begin{align}
\langle V_{1}(y_{1})V_{2}(y_{2})\int_{y_{2}}^{y_{3}}dyU_{3}(y)&\left[\int d\bar{z}_{4}V(\theta)\bar{U}(\bar{\theta})e^{ikX}+\int dz_{4}U(\theta)\bar{V}(\bar{\theta})e^{ikX}\right]\times\nonumber \\
&\int d^{2}z_{5}U_{5}(z_{5},\bar{z}_{5})...\int d^{2}z_{N}U_{N}(z_{N},\bar{z}_{N})\rangle.
\end{align}
From the above result we can see the fact that we can fix only ``one
half" of the closed string. Now we can perform the same steps for
another open string. For each of the above terms we obtain 
\begin{align*}
\langle V_{1}(y_{1})&\int_{y_{1}}^{y_{2}}dy^{\prime}\int d\bar{z}_{4}[Q,U_{2}(y^{\prime})]\{\Theta(y^{\prime}-y_{2})-\Theta(y^{\prime}-y_{1})+\ln(\left|z_{3}-z_{4}\right|^{2})\}\times\\
&V(\theta)\bar{U}(\bar{\theta})e^{ikX}\int_{y_{2}}^{y_{3}}dyU_{3}(y)\int d^{2}z_{5}U_{5}(z_{5},\bar{z}_{5})...\int d^{2}z_{N}U_{N}(z_{N},\bar{z}_{N})\rangle.
\end{align*}
 and using the fact that $Q^{2}=0$ we get, after pulling of the BRST
operator 
\begin{align*}
\langle V_{1}(y_{1})&\int_{y_{1}}^{y_{2}}dy^{\prime}U_{2}(y^{\prime})\int d\bar{z}_{4}\{\Theta(y^{\prime}-y_{2})-\Theta(y^{\prime}-y_{1})+\ln(\left|z_{3}-z_{4}\right|^{2})\}V(\theta)\bar{\partial}(\bar{V}(\bar{\theta})e^{ikX})\\
&\int_{y_{2}}^{y_{3}}dyU_{3}(y)\int d^{2}z_{5}U_{5}(z_{5},\bar{z}_{5})...\int d^{2}z_{N}U_{N}(z_{N},\bar{z}_{N})\rangle.
\end{align*}
Using the same arguments as before we finally get 
\begin{eqnarray}
\langle V_{1}(y_{1})\int_{y_{1}}^{y_{2}}dy^{\prime}U_{2}(y^{\prime})\int_{y_{2}}^{y_{3}}dyU_{3}(y)V(\theta)\bar{V}(\bar{\theta})e^{ikX}(z_{3},\bar{z}_{3})\times\nonumber \\\int d^{2}z_{5}U_{5}(z_{5},\bar{z}_{5})...\int d^{2}z_{N}U_{N}(z_{N},\bar{z}_{N})\rangle.
\end{eqnarray}
and we end our proof. This symmetry will be used to stablish the equivalence with RNS and to simplify the computation of the three point tree level Type I amplitude.

\section{Equivalence for Amplitudes involving four fermions}

In this section we show the equivalence for amplitudes involving
four fermions. Here and in the next two sections we choose the
measure constant $c=2880$ in order to compare the results with the RNS formalism.
In another section we must choose $c=1$ for comparison with previous
computations of the three point function. The general amplitude is
given by 
\begin{eqnarray*}
\mathcal{A} & = & \left\langle VVV\int U...\int U\right\rangle 
\end{eqnarray*}
The case with four open or two closed string fermionic states would
reduce to the case already studied at \cite{Berkovits:2000ph}. Therefore we
must consider the case with one fermionic closed and two fermionic
open strings. With the help of cyclic symmetry we can choose the closed and one open string fixed. Therefore the amplitude can be written as

\begin{eqnarray}
\mathcal{A}=\langle\xi_{1}^{\alpha}\widetilde{\xi}^{\beta}V_{\alpha}^{F}(z_{1})\bar{V}{}_{\beta}^{F}(\bar{z}_{1})\xi_{2}^{\gamma}V_{\gamma}^{F}(y_{2})\int dy_{3}\xi_{3}^{\delta}p_{\delta}\times \nonumber\\
\int d^{2}z_{4}h^{m}\widetilde{h}^{n}U_{m}^{B}(z_{4})\bar{U}_{n}^{B}(\bar{z}_{4})\int dy_{5}a^{p}U_{p}...\int dy_{n}a^{q}U_{q}e^{i\sum_{r=1}^{N}k_{r}\cdot x(z_{r})}\rangle,
\end{eqnarray}
where the dots mean an arbitrary number of closed or open bosonic
strings. Now we must use the explicit shape of the vertex operators
and look for the terms with five thetas. For this, note that $p_{\delta}$
acts as a theta derivative. Because the fixed fermionic operators
contribute with at least six thetas, all the integrated bosonic operators
must contribute with zero thetas. With this, the amplitude
is given by

\begin{eqnarray}
\mathcal{A} & = & -\frac{\alpha^{\prime}}{2\times27}\langle\xi_{1}^{\alpha}\widetilde{\xi}^{\beta}f_{\alpha}(z_{1})\bar{f}{}_{\beta}(\bar{z}_{1})\xi_{2}^{\gamma}f_{\gamma}(y_{2})\int dy_{3}\xi_{3}^{\delta}p_{\delta}\times \nonumber \\
&& \int d^{2}z_{4}h_{4}^{m_{4}}\widetilde{h}_{4}^{n_{4}}(\partial x_{m_{4}}-ik^{p_{4}}M_{m_{4}p_{4}})(\bar{\partial}x_{n_{4}}-ik^{q_{4}}M_{n_{4}q_{4}})\times  \\
&&\int dy_{5}a_{5}^{m_{5}}(\partial x_{m_{5}}-ik^{p_{5}}M_{m_{5}p_{5}})...\int dy_{N}a_{N}^{m_{n}}(\partial x_{m_{n}}-ik^{p_{n}}M_{m_{n}p_{n}})e^{i\sum_{r=1}^{N}k_{r}\cdot x(z_{r})}\rangle.\nonumber
\end{eqnarray}
In the RNS case the amplitude is given by \cite{Friedan:1985ge}

\begin{eqnarray}
{\cal A}_{RNS}&=&-\langle\xi_{1}^{\alpha}ce^{-\frac{\phi}{2}}\Sigma_{\alpha}(z_{1})\widetilde{\xi}_{1}^{\beta}\bar{c}e^{-\frac{\bar{\phi}}{2}}\bar{\Sigma}{}_{\beta}(\bar{z}_{1})\xi_{3}^{\gamma}ce^{-\frac{\phi}{2}}\Sigma_{\gamma}(y_{2})\int dy_{3}\xi_{3}^{\delta}e^{-\frac{\phi}{2}}\Sigma_{\delta}(y_{3})\nonumber \\ 
&&\int d^{2}z_{4}h_{4}^{m_{4}}\widetilde{h}_{4}^{n_{4}}(\bar{\partial}x_{m_{4}}(z_{4})-i\frac{\alpha^{\prime}}{2}k_{4}^{p_{4}}\bar{\psi}_{m_{4}}\bar{\psi}_{p_{4}}(z_{4}))(\partial x_{n_{4}}(z_{4})-i\frac{\alpha^{\prime}}{2}k_{4}^{q_{4}}\psi_{n_{4}}\psi_{q_{4}}(z_{4}))\times \nonumber \\ 
&&\int dy_{6}a_{5}^{n_{5}}(\partial x_{n_{5}}(y_{5})-i\frac{\alpha^{\prime}}{2}k_{5}^{q_{5}}\psi_{n_{5}}\psi_{q_{5}}(z_{5}))...\nonumber \\
&&\times\int dz_{N}a_{N}^{n}(\partial x_{n}(z_{N})-i\frac{\alpha^{\prime}}{2}k_{N}^{q}\psi_{n}\psi_{q}(z_{N}))e^{i\sum_{r=1}^{N}k_{r}\cdot x(z_{r})}\rangle
\end{eqnarray}
and all objects above have been defined before. The OPEs between the
$x$ fields are the same in both formalisms, therefore we just need
to prove the equivalence for the part independent of $x$ . It has
been argued before in \cite{Berkovits:2000ph} that all the OPEs between the bosonic
integrated operators and the fermionic operators are the same in both
formalisms. This is also valid here when we consider the OPEs given
in section $2$ and the fact that here we have a mixing between left
and right movers. This guarantee that the dependence on $z_{4}...z_{N}$
are the same. In order to prove the equivalence we just have to show that

\begin{eqnarray}
\frac{1}{27}\langle f_{\alpha}(z_{1})\bar{f}_{\beta}(\bar{z}_{1})f_{\gamma}(y_{2})p_{\delta}(y_{3})\rangle=\langle ce^{-\frac{\phi}{2}}\Sigma_{\alpha}(z_{1})\bar{c}e^{-\frac{\bar{\phi}}{2}}\bar{\Sigma}{}_{\beta}(\bar{z}_{1})ce^{-\frac{\phi}{2}}\Sigma_{\gamma}(y_{2})e^{-\frac{\phi}{2}}\Sigma_{\delta}(y_{3})\rangle.
\end{eqnarray}

The right hand side can be computed easily and results in\cite{Friedan:1985ge}

\begin{equation}
\langle ce^{-\frac{\phi}{2}}\Sigma_{\alpha}(z_{1})\bar{c}e^{-\frac{\bar{\phi}}{2}}\bar{\Sigma}{}_{\beta}(\bar{z}_{1})ce^{-\frac{\phi}{2}}\Sigma_{\gamma}(y_{2})e^{-\frac{\phi}{2}}\Sigma_{\delta}(y_{3})\rangle=\{\frac{\gamma_{\alpha\delta}^{m}\gamma_{m\,\beta\gamma}}{z_{1}-y_{3}}+\frac{\gamma_{\beta\delta}^{m}\gamma_{m\,\gamma\alpha}}{\bar{z}_{1}-y_{3}}+\frac{\gamma_{\gamma\delta}^{m}\gamma_{m\,\alpha\beta}}{y_{2}-y_{3}}\}.
\end{equation}
For the left hand side, following the same lines of reasoning
of \cite{Berkovits:2000ph}, we must analyze the poles of $p_{\delta}(y_{3})$. Here
we must be careful with the mixing of left an right movers. That is
why, beyond the pole coming from $y_{3}\to z_{1}$ we also have a
pole in $y_{3}\to\bar{z}_{1}$. Therefore, in the Type I case we have
the following residue

\begin{eqnarray}
&&\frac{1}{27}\langle\{[(\gamma_{\alpha\delta}^{m}(\lambda\gamma_{m}\theta)(z_{1})-(\gamma_{m}\lambda)_{\delta}(\gamma^{m}\theta)_{\alpha}(z_{1})](\bar{\lambda}\gamma^{n}\bar{\theta})(\gamma_{n}\bar{\theta})_{\beta}(\bar{z}_{1})+ \\
&&+(\lambda\gamma_{m}\theta)(\gamma^{m}\theta)_{\alpha}(z_{1})[\gamma_{\delta\beta}^{n}(\bar{\lambda}\gamma_{n}\bar{\theta})(\bar{z}_{1})-(\bar{\lambda}\gamma^{n})_{\delta}(\gamma_{n}\bar{\theta})_{\beta}(\bar{z}_{1})]\}(\lambda\gamma^{p}\theta)(\gamma_{p}\theta)_{\gamma}(y_{2})\rangle.\nonumber
\end{eqnarray}

The above expression can be simplified if we use
\begin{eqnarray}
&&-(\gamma_{m}\lambda)_{\delta}(\gamma^{m}\theta)_{\alpha}(z_{1})(\bar{\lambda}\gamma^{n}\bar{\theta})(\gamma_{n}\bar{\theta})_{\beta}(\bar{z}_{1})-(\lambda\gamma_{m}\theta)(\gamma^{m}\theta)_{\alpha}(z_{1})(\bar{\lambda}\gamma^{n})_{\delta}(\gamma_{n}\bar{\theta})_{\beta}(\bar{z}_{1})\nonumber \\
&=&\lambda D\{\frac{1}{2}(\gamma^{m}\theta)_{\alpha}(\gamma_{m}\theta)_{\delta}(\bar{\lambda}\gamma^{n}\bar{\theta})(\gamma_{n}\bar{\theta})_{\beta}\}+\bar{\lambda}\bar{D}\{\frac{1}{2}(\lambda\gamma_{m}\theta)(\gamma^{m}\theta)_{\alpha}(\gamma^{m}\bar{\theta})_{\beta}(\gamma_{m}\bar{\theta})_{\delta}\}+\nonumber \\
&&+\frac{1}{2}\gamma_{\alpha\delta}^{m}(\lambda\gamma_{m}\theta)(\bar{\lambda}\gamma^{n}\bar{\theta})(\gamma_{n}\bar{\theta})_{\beta}+\frac{1}{2}(\lambda\gamma_{m}\theta)(\gamma^{m}\theta)_{\alpha}\gamma_{\alpha\delta}^{m}(\bar{\lambda}\gamma_{m}\bar{\theta})
\end{eqnarray}
and we obtain that the lhs is given by
\begin{eqnarray}
&&\frac{2}{\alpha^{\prime}}Q_{o}\{\frac{1}{2}(\gamma^{m}\theta)_{\alpha}(\gamma_{m}\theta)_{\delta}(\bar{\lambda}\gamma^{n}\bar{\theta})(\gamma_{n}\bar{\theta})_{\beta}+\frac{1}{2}(\lambda\gamma_{m}\theta)(\gamma^{m}\theta)_{\alpha}(\gamma^{m}\bar{\theta})_{\beta}(\gamma_{m}\bar{\theta})_{\delta}\}+\nonumber \\
&&+\frac{1}{2}\gamma_{\alpha\delta}^{m}(\lambda\gamma_{m}\theta)(\bar{\lambda}\gamma^{n}\bar{\theta})(\gamma_{n}\bar{\theta})_{\beta}+\frac{1}{2}(\lambda\gamma_{m}\theta)(\gamma^{m}\theta)_{\alpha}\gamma_{\alpha\delta}^{m}(\bar{\lambda}\gamma_{m}\bar{\theta}).
\end{eqnarray}

Due to the BRST invariance of the amplitude we can always exchange
\begin{eqnarray}
&&-(\gamma_{m}\lambda)_{\delta}(\gamma^{m}\theta)_{\alpha}(z_{1})(\bar{\lambda}\gamma^{n}\bar{\theta})(\gamma_{n}\bar{\theta})_{\beta}(\bar{z}_{1})-(\lambda\gamma_{m}\theta)(\gamma^{m}\theta)_{\alpha}(z_{1})(\bar{\lambda}\gamma^{n})_{\delta}(\gamma_{n}\bar{\theta})_{\beta}(\bar{z}_{1})\nonumber \\ &&\rightarrow\frac{1}{2}\gamma_{\alpha\delta}^{m}(\lambda\gamma_{m}\theta)(\bar{\lambda}\gamma^{n}\bar{\theta})(\gamma_{n}\bar{\theta})_{\beta}+\frac{1}{2}(\lambda\gamma_{m}\theta)(\gamma^{m}\theta)_{\alpha}\gamma_{\alpha\delta}^{m}(\bar{\lambda}\gamma_{m}\bar{\theta}).
\end{eqnarray}

Using the above result our expression above is simplified to give

\begin{eqnarray}
&&\frac{\gamma_{\alpha\delta}^{n}}{18}\langle(\lambda\gamma_{n}\theta)(z_{1})(\bar{\lambda}\gamma^{m}\bar{\theta})(\gamma_{m}\bar{\theta})_{\beta}(\bar{z}_{1})(\lambda\gamma^{p}\theta)(\gamma_{p}\theta)_{\gamma}(y_{2})\rangle \nonumber\\ 
&&+\frac{\gamma_{\delta\beta}^{n}}{18}\langle(\lambda\gamma_{m}\theta)(\gamma^{m}\theta)_{\alpha}(z_{1})(\bar{\lambda}\gamma_{n}\bar{\theta})(\bar{z}_{1})(\lambda\gamma^{p}\theta)(\gamma_{p}\theta)_{\gamma}(y_{2})\rangle \nonumber \\
&&\equiv \frac{\gamma_{\alpha\delta}^{n}}{18}(G_{n})_{\beta\gamma}+\frac{\gamma_{\delta\beta}^{n}}{18}(H_{n})_{\alpha\gamma}
\end{eqnarray}
and each of the above terms is easily computed. First of all the quantity
in brackets is Lorentz invariant and therefore must be proportional
to $(\gamma_{n})_{\beta\gamma}$. If we use our measure we finally
obtain $(G_{n})_{\beta\gamma}=18(\gamma_{n})_{\beta\gamma}$ and $(H_{n})_{\alpha\gamma}=18(\gamma_{n})_{\alpha\gamma}$.
Therefore the residue when $y_{3}\to z_{1}$ is the same as in the
RNS case. The argument for the $y_{2}$ pole is identical to that
with only open string and is unnecessary here. The above result can
be summarized in the identity 
\begin{eqnarray}
\frac{1}{27}\langle f_{\alpha}(z_{1})f_{\beta}(\bar{z}_{1})f_{\gamma}(y_{2})p_{\delta}(y_{3})\rangle=\{\frac{\gamma_{\alpha\delta}^{m}\gamma_{m\,\beta\gamma}}{z_{1}-y_{3}}+\frac{\gamma_{\beta\delta}^{m}\gamma_{m\,\gamma\alpha}}{\bar{z}_{1}-y_{3}}+\frac{\gamma_{\gamma\delta}^{m}\gamma_{m\,\alpha\beta}}{y_{2}-y_{3}}\}.
\end{eqnarray}
In this way, we have shown the equivalence with the RNS formalism for
amplitudes involving four fermions. Now we must consider the case involving 
two fermions.

\section{Equivalence for Amplitudes Involving Two Fermions}

As before, the general amplitude is given by
\begin{eqnarray*}
\mathcal{A} & = & \left\langle VVV\int U...\int U\right\rangle .
\end{eqnarray*}
The case with two open or one closed string fermionic states would be
reduced to the case already studied at \cite{Berkovits:2000ph}. Therefore we
must consider the case with one fermionic closed string, which involves
one fermion for each moving sector. With the cyclic symmetry for Type I
strings we can choose the closed and one open string fixed. This will give the contribution of four thetas. The other unintegrated vertex operator is bosonic and therefore we
get one more theta, which fullfills five thetas. The last vertex operator
is bosonic and integrated and therefore contributes with zero thetas
as expected. Therefore, the amplitude can be written as
\begin{eqnarray*}
\mathcal{A} & = & \langle\xi_{1}^{\alpha}\widetilde{\xi}^{\beta}V_{\alpha}^{F}(z_{1})\bar{V}{}_{\beta}^{F}(\bar{z}_{1})b_{2}^{m_{2}}V_{m_{2}}^{B}(y_{2})\int dy_{3}a^{m_{3}}U_{m_{3}}(y_{3})\int d^{2}z_{4}h^{m}\widetilde{h}^{n}U_{m}^{B}(z_{4})\bar{U}_{n}^{B}(\bar{z}_{4})\times \\
&&\int dy_{5}a^{p}U_{p}...\int dy_{n}a^{q}U_{q}e^{i\sum_{r=1}^{N}k_{r}\cdot x(z_{r})}\rangle,
\end{eqnarray*}
where the dots mean an arbitrary number of closed or open bosonic
strings. Now we must use the explicit shape of the vertex operators
and look for the terms with five thetas. To achieve this, note that
the fixed operators already contribute with at last five thetas. Therefore,
all the integrated bosonic operators must contribute with zero thetas.
With this, the amplitude is given by
\begin{eqnarray*}
&&\mathcal{A}=-\frac{\alpha^{\prime}}{18}\langle\xi_{1}^{\alpha}\widetilde{\xi}^{\beta}f_{\alpha}(z_{1})\bar{f}{}_{\beta}(\bar{z}_{1})a_{2}^{m_{2}}b_{m_{2}}(y_{2})\int dy_{3}a^{m_{3}}(\partial x_{m_{3}}-ik^{p_{3}}M_{m_{3}p_{3}})(y_{3})\times \\
 &  & \int d^{2}z_{4}h_{4}^{m_{4}}\widetilde{h}_{4}^{n_{4}}(\partial x_{m_{4}}-ik^{p_{4}}M_{m_{4}p_{4}})(\bar{\partial}x_{n_{4}}-ik^{q_{4}}M_{n_{4}q_{4}})\int dy_{5}a_{5}^{m_{5}}(\partial x_{m_{5}}-ik^{p_{5}}M_{m_{5}p_{5}})...\\
 &&\int dy_{N}a_{N}^{m_{n}}(\partial x_{m_{n}}-ik^{p_{n}}M_{m_{n}p_{n}})e^{i\sum_{r=1}^{N}k_{r}\cdot x(z_{r})}\rangle.
\end{eqnarray*}
In the RNS case the amplitude is given by

\begin{eqnarray}
&&{\cal A}_{RNS}=-\langle\xi_{1}^{\alpha}ce^{-\frac{\phi}{2}}\Sigma_{\alpha}(z_{1})\widetilde{\xi}_{1}^{\beta}\bar{c}e^{-\frac{\bar{\phi}}{2}}\bar{\Sigma}{}_{\beta}(\bar{z}_{1})a_{2}^{n_{2}}ce^{-\phi}\psi_{n_{2}}(y_{2})\nonumber \times \\
&&\int dy_{3}a_{3}^{n_{3}}(\partial x_{n_{3}}(z_{3})-i\frac{\alpha^{\prime}}{2}k_{3}^{q_{3}}\psi_{n_{3}}\psi_{q_{3}}(y_{3}))\times \nonumber \\
&&\int d^{2}z_{4}h_{4}^{m_{4}}\widetilde{h}_{4}^{n_{4}}(\bar{\partial}x_{m_{4}}(z_{4})-i\frac{\alpha^{\prime}}{2}k_{4}^{p_{4}}\bar{\psi}_{m_{4}}\bar{\psi}_{p_{4}}(z_{4}))(\partial x_{n_{4}}(z_{4})-i\frac{\alpha^{\prime}}{2}k_{4}^{q_{4}}\psi_{n_{4}}\psi_{q_{4}}(z_{4}))\times \nonumber \\
&&\int dy_{6}a_{5}^{n_{5}}(\partial x_{n_{5}}(y_{5})-i\frac{\alpha^{\prime}}{2}k_{5}^{q_{5}}\psi_{n_{5}}\psi_{q_{5}}(z_{5}))\times ... \\
&&\int dz_{N}a_{N}^{n}(\partial x_{n}(z_{N})-i\frac{\alpha^{\prime}}{2}k_{N}^{q}\psi_{n}\psi_{q}(z_{N}))e^{i\sum_{r=1}^{N}k_{r}\cdot x(z_{r})}\rangle.\nonumber
\end{eqnarray}

The OPEs between the $x$ fields are again the same in both formalisms,
therefore we just need to prove equivalence for the $x$ independent
part. Like in the last section, we use the fact that it has been
argued before in \cite{Berkovits:2000ph} that all the OPEs between the bosonic integrated
operators and the fermionic operators are the same in both formalisms.
This is also valid here when we consider that all the OPEs, in both
formalisms, have a mixing between left and right movers. This guarantee
that the dependence on $z_{4}...z_{N}$ are the same, and to prove
the equivalence we just need to show that

\begin{equation}
\frac{1}{18}\langle f_{\alpha}(z_{1})\bar{f}_{\beta}(\bar{z}_{1})b_{m}(z_{3})\rangle=\langle ce^{-\frac{\phi}{2}}\Sigma_{\alpha}(z_{1})\bar{c}e^{-\frac{\bar{\phi}}{2}}\bar{\Sigma}{}_{\beta}(\bar{z}_{1})ce^{-\phi}\psi_{m}(y_{2})\rangle.
\end{equation}

The RNS result for the right hand side is given by $(\gamma_{m})_{\alpha\beta}$.
The left hand side is given by
\begin{equation}
\frac{1}{18}\langle(\lambda\gamma^{p}\theta)(\gamma_{p}\theta)_{\gamma}(z_{1})(\bar{\lambda}\gamma^{m}\bar{\theta})(\gamma_{m}\bar{\theta})_{\beta}(\bar{z}_{1})(\lambda\gamma_{n}\theta)(y_{2})\rangle=\frac{1}{18}(G_{n})_{\beta\gamma}=(\gamma_{n})_{\beta\gamma}
\end{equation}
and therefore we have shown the equivalence between the amplitudes. Next section we consider the last case, that is, amplitudes with zero fermions.

\section{Equivalence for Amplitudes Involving Zero Fermions}

Now we arrive to the last and by far the more involved point. The case with only
open or only closed string bosonic states would also be reduced to the
case already studied at \cite{Berkovits:2000ph}. Therefore we must consider
the case with at least one closed and one open string. With the cyclic
symmetry for Type I strings we can choose the closed and one open string fixed. Therefore, the amplitude can be written as

\begin{eqnarray*}
\mathcal{A} & = & \langle h^{m_{1}}\widetilde{h}^{n_{1}}V_{m_{1}n_{1}}^{B}(z_{1},\bar{z}_{1})b_{2}^{m_{2}}V_{m_{2}}^{B}(y_{2})\int dy_{3}a^{m_{3}}U_{m_{3}}(y_{3})\int d^{2}z_{4}h^{m_{4}}\widetilde{h}^{n_{4}}U_{m_{4}}^{B}(z_{4})\bar{U}_{n_{4}}^{B}(\bar{z}_{4})\times \\
&&\int dy_{5}a_{5}^{m_{5}}U_{m_{5}}...\int dy_{n}a_{n}^{m_{n}}U_{m_{n}}\rangle.
\end{eqnarray*}

The strategy in this case is to use supersymmetry to simplify the
computations. The supersymmetry transformations are given by
\begin{eqnarray*}
\{q_{\alpha},V_{m}^{B}\}=\frac{i}{2}k^{n}(\gamma_{mn})_{\alpha}{}^{\beta}V_{\beta}^{F}+Q(\Omega_{m\alpha}),
\end{eqnarray*}

\begin{eqnarray*}
[q_{\alpha},U_{m}^{B}]=\frac{i}{2}k^{n}(\gamma_{mn})_{\alpha}{}^{\beta}U_{\beta}^{F}-\partial(\Omega_{m\alpha}),\quad\{q_{\alpha},U_{\beta}^{F}\}=\gamma_{\alpha\beta}^{m}U_{m}^{B}+\partial(\Sigma_{\alpha\beta}),
\end{eqnarray*}

\begin{eqnarray*}
[q_{\alpha},V_{m_1n_1}^{B}]=\frac{i}{2}k^{q}(\gamma_{m_1q})_{\alpha}{}^{\beta}V_{\beta n_1}^{F}+\frac{i}{2}k^{q}(\gamma_{n_1q})_{\alpha}{}^{\beta}V_{m_1\beta}^{F}+Q(\Omega_{m_1n_1\alpha}),
\end{eqnarray*}

\begin{equation}
 \quad[q_{\alpha},V_{\beta n_1}^{F}]=\gamma_{\alpha\beta}^{m_1}V_{m_1n_1}^{B}+\frac{i}{2}k^{q}(\gamma_{n_1q})_{\alpha}{}^{\delta}V_{\alpha \delta}^{F}+Q(\Sigma_{\alpha\beta n_1}),
\end{equation}
we can get
\begin{equation}
V_{m_1n_1}^{B}=\frac{1}{16} \gamma^{\alpha\beta}_{m_1}[q_{\alpha},V_{\beta n_1}^{F}]-\frac{i}{32}\gamma^{\alpha\beta}_{m_1}k^{q}(\gamma_{n_1q})_{\alpha}{}^{\delta}V_{\alpha \delta}^{F}+Q(\Sigma_{\alpha\beta n_1}).
\end{equation}

The above expression can be used in the amplitude to replace the fixed closed string. Let us analyze the second term of the right side, which is more simple. After using the argument that the OPEs between the $x$ field is the same for the 
RNS and PS formalism, we must prove that the part independent of $x$ is equivalent. However, note that in this case we can factorize the left and right movers of $V^B_{\alpha\delta}(z,\bar{z})=V^F_{\alpha}(z)\bar{V}^F_{\delta}(\bar{z})$ and we get the same case as that with 
two fermions, but with a different polarization, namely
\begin{eqnarray*}
\mathcal{A}_2 &=& -\frac{i}{32}\langle h^{m_{1}}\widetilde{h}^{n_{1}}k_{1}^{p_{1}}(\gamma_{m_{1}p_{1}})_{\alpha}{}^{\sigma}V_{\sigma}^{F}(z_{1})\gamma_{n_{1}}^{\alpha\beta}\bar{V}_{\beta}^{F}(\bar{z}_{1})b_{2}^{m_{2}}V_{m_{2}}^{B}(y_{2})\int dy_{3}a^{m_{3}}U_{m_{3}}(y_{3})\times \\
 &&\int d^{2}z_{4}h^{m_{4}}\widetilde{h}^{n_{4}}U_{m_{4}}^{B}(z_{4})\bar{U}_{n_{4}}^{B}(\bar{z}_{4})\int dy_{5}a_{5}^{m_{5}}U_{m_{5}}...\int dy_{n}a_{n}^{m_{n}}U_{m_{n}}\rangle.
\end{eqnarray*}
For the first term we use the fact that the amplitude is supersymmetric, what will
give rise to many terms. We also use again the fact that we just need to consider the part independent of $x$ to use $V_{\beta n_1}^{F}=\bar{V}^{F}_\beta (\bar{z}){V}^{B}_{n_1} (z)$. For example, 
when the operator acts in $y_{2}$ we get

\begin{eqnarray*}
\mathcal{A}_1 & = & -\frac{1}{16}\langle h^{m_{1}}\widetilde{h}^{n_{1}}V_{n_{1}}^{B}(z_{1})\gamma_{m_{1}}^{\alpha\beta}V_{\beta}^{F}(\bar{z}_{1})b_{2}^{m_{2}}\{q_{\alpha},V_{m_{2}}^{B}(y_{2})\}\int dy_{3}a^{m_{3}}U_{m_{3}}(y_{3})\times \\
&&\int d^{2}z_{4}h^{m_{4}}\widetilde{h}^{n_{4}}U_{m_{4}}^{B}(z_{4})\bar{U}_{n_{4}}^{B}(\bar{z}_{4})\int dy_{5}a_{5}^{m_{5}}U_{m_{5}}...\int dy_{n}a_{n}^{m_{n}}U_{m_{n}}\rangle,
\end{eqnarray*}
and using again the previous supersymmetry transformation the expression above turns to be the same as the one used to compute the amplitude with two fermions. They only differ in the polarizations. When $q_{\alpha}$ acts in an integrated
operator, $y_{3}$ for example, we get 

\begin{eqnarray*}
\mathcal{A} & = & -\frac{1}{16}\langle h^{m_{1}}\widetilde{h}^{n_{1}}V_{m_{1}}^{B}(z_{1})\gamma_{n_{1}}^{\alpha\beta}V_{\beta}^{F}(\bar{z}_{1})b_{2}^{m_{2}}V_{m_{2}}^{B}(y_{2})\int dy_{3}a^{m_{3}}\{q_{\alpha},U_{m_{3}}^{B}(y_{3})\} \nonumber \times \\ 
&&\int d^{2}z_{4}h^{m_{4}}\widetilde{h}^{n_{4}}U_{m_{4}}^{B}(z_{4})\bar{U}_{n_{4}}^{B}(\bar{z}_{4})\int dy_{5}a_{5}^{m_{5}}U_{m_{5}}...\int dy_{n}a_{n}^{m_{n}}U_{m_{n}}\rangle,
\end{eqnarray*}
that give us
\begin{eqnarray*}
\mathcal{A} & = & -\frac{1}{16}\langle h^{m_{1}}\widetilde{h}^{n_{1}}V_{m_{1}}^{B}(z_{1})\gamma_{n_{1}}^{\alpha\beta}V_{\beta}^{F}(\bar{z}_{1})b_{2}^{m_{2}}V_{m_{2}}^{B}(y_{2})\int dy_{3}a^{m_{3}}\frac{i}{2}k_{3}^{p_{3}}(\gamma_{m_{3}p_{3}})_{\alpha}{}^{\sigma}U_{\sigma}^{F}(y_{3})\times \\
 &&\int d^{2}z_{4}h^{m_{4}}\widetilde{h}^{n_{4}}U_{m_{4}}^{B}(z_{4})\bar{U}_{n_{4}}^{B}(\bar{z}_{4})\int dy_{5}a_{5}^{m_{5}}U_{m_{5}}...\int dy_{n}a_{n}^{m_{n}}U_{m_{n}}\rangle.
\end{eqnarray*}

Using now the cyclic symmetry we can
exchange $V_{m_{2}}^{B}(y_{2})\int dy_{3}U_{\sigma}^{F}(y_{3})$ for
$\int dy_{3}V_{m_{2}}^{B}(y_{3})V_{\sigma}^{F}(y_{2})$. Then again
this amplitude is related to the case with two fermions of the last
section. Therefore, any pure spinor amplitude is related to a combination
of RNS amplitudes involving $N-2$ bosons. We should show here that
this amplitude is related to the $N$ bosons amplitude in this formalism.
At this point is must be clear that the argument used in \cite{Berkovits:2000ph} is also valid here.  
We must focus on an application of these ideas to simplify the tree level three point amplitude in 
the next section.

\section{Simplifying the Three Point Type I Amplitude}

In this section we must use the ideas of the last ones to simplify
the computation of the three point amplitude computed in \cite{Alencar:2008fy}.
With this technique we will see that no computer program will be needed
and so we gain a considerable simplification. For the comparison with
the previous result we must choose the measure constant $c=1$. With this, the identities of the last sections are
modified to

\begin{eqnarray}
\frac{1}{27}\langle f_{\alpha}(z_{1})f_{\beta}(\bar{z}_{1})f_{\gamma}(y_{2})p_{\delta}(y_{3})\rangle=\frac{1}{2880}\{\frac{\gamma_{\alpha\delta}^{m}\gamma_{m\,\beta\gamma}}{z_{1}-y_{3}}+\frac{\gamma_{\beta\delta}^{m}\gamma_{m\,\gamma\alpha}}{\bar{z}_{1}-y_{3}}+\frac{\gamma_{\gamma\delta}^{m}\gamma_{m\,\alpha\beta}}{y_{2}-y_{3}}\}
\end{eqnarray}
and

\begin{equation}
\frac{1}{18}\langle f_{\alpha}(z_{1})\bar{f}_{\beta}(\bar{z}_{1})b_{m}(z_{3})\rangle=\frac{(\gamma_{n})_{\beta\gamma}}{2880}
\end{equation}
which we must use throughout this section.

\subsection{Kalb-Ramond and Two Photinos}

In this case we will always have a fermion vertex operator integrated,
but using the above identity the result will be obtained directly.
In Type I superstring fermion-fermion contribution will give us the
Kalb-Ramond field. Therefore this will give us the amplitude for one
Kalb-Ramond and two Photinos. This is the simplest amplitude, and
using our previous definitions this is given by 
\begin{eqnarray*}
\mathcal{A} & = & -\frac{\alpha^{\prime}}{2\times27}\langle\xi^{\alpha}f_{\alpha}\widetilde{\xi}^{\beta}f_{\beta}\xi_{2}^{\gamma}f_{\gamma}\int dy\xi_{3}^{\delta}p_{\delta}\rangle.
\end{eqnarray*}

Fixing now the position of the operators, the above amplitude can
be written as 

\begin{eqnarray*}
\mathcal{A} & = & -\frac{\alpha^{\prime}}{2\times27}\xi^{\alpha}\widetilde{\xi}^{\beta}\xi_{2}^{\gamma}\xi_{3}^{\delta}\int dy\langle f_{\alpha}(ia)f_{\beta}(-ia)f_{\gamma}(\infty)p_{\delta}(y)\rangle
\end{eqnarray*}
 and from the above identity we get

\begin{eqnarray*}
\mathcal{A} & = & -\frac{\alpha^{\prime}}{2\times2880}\xi^{\alpha}\widetilde{\xi}^{\beta}\xi_{2}^{\gamma}\xi_{3}^{\delta}\int dy\left[\frac{\gamma_{\alpha\delta}^{m}\gamma_{m\,\beta\gamma}}{ia-y}+\frac{\gamma_{\beta\delta}^{m}\gamma_{m\,\gamma\alpha}}{-ia-y}\right].
\end{eqnarray*}
Using residues we arrive at

\begin{eqnarray*}
\mathcal{A} & = & \frac{2\alpha^{\prime}}{2880}\pi i\xi^{\alpha}\widetilde{\xi}^{\beta}\xi_{2}^{\gamma}\xi_{3}^{\delta}\gamma_{\alpha\delta}^{m}\gamma_{m\,\beta\gamma},
\end{eqnarray*}
and using the fact that, in type I superstring

\[
\xi^{\alpha}\widetilde{\xi}^{\beta}=\frac{1}{96}\gamma_{abc}^{\alpha\beta}H^{abc}
\]
we obtain

\begin{eqnarray*}
\mathcal{A} & = & \frac{i\pi\alpha^{\prime}}{48\times2880}\gamma_{abc}^{\alpha\beta}H^{abc}\xi_{3}^{\gamma}\xi_{4}^{\delta}\gamma_{\alpha\delta}^{m}\gamma_{m\,\beta\gamma}.
\end{eqnarray*}
Using now the identity

\begin{eqnarray*}
\gamma^{a}\gamma^{bcd}\gamma_{a} & = & -4\gamma^{bcd}
\end{eqnarray*}
we obtain

\begin{eqnarray*}
\mathcal{A} & = & -\frac{i\pi\alpha^{\prime}}{720\times48}H^{abc}\xi_{3}\gamma_{abc}^{\alpha\beta}\xi_{4},
\end{eqnarray*}
which agrees with the previous computation for the Kalb-Ramond and
two photinos with pure spinor formalism \cite{Alencar:2008fy}.

\subsection{Two Fermions}

In this subsection we will consider all the possibilities that contain
two fermions. These are: Kalb-Ramond and two photons; one gravitino/dilatino,
one photon and one photino; one graviton/dilaton and two photinos.

\subsubsection{Kalb Ramond and two Photons}

In this case we have

\begin{eqnarray*}
\mathcal{A} & = & \frac{1}{18}\xi^{\alpha}\tilde{\xi}^{\beta}a_{2}^{n}a_{3}^{m}\int dy\langle f_{\alpha}(ia)f_{\beta}(-ia)b_{n}^{2}(\infty)[\dot{x}_{m}-ik_{3}^{p}M_{mp}]\rangle
\end{eqnarray*}
and here there is one thing that must be observed. Note that when
we contract $\dot{x}$ of the integrated open string with the unintegrated
closed string we get a null result because

\begin{eqnarray*}
\int dy\langle f_{\alpha}(ia)f_{\beta}(-ia)b_{n}^{2}(\infty)\dot{x}_{m}\rangle=\int dy[-\frac{i\alpha^{\prime}k_{m}^{1}}{(y-ia)}-\frac{i\alpha^{\prime}k_{m}^{1}}{(y+ia)}]\langle f_{\alpha}(ia)f_{\beta}(-ia)b_{n}^{2}(\infty)\rangle & =0.
\end{eqnarray*}
and this will be used throughout this whole section. Therefore in
the above expression we are left only with the second term and using
the OPEs we obtain

\begin{eqnarray*}
\mathcal{A} & = & \frac{1}{18}\xi^{\alpha}\tilde{\xi}^{\beta}a_{2}^{n}a_{3}^{m}\times \nonumber \\ &&\int dy\langle[-\frac{i\alpha^{\prime}}{4(y-ia)}k_{3}^{p}(\gamma_{mp})_{\alpha}^{\,\sigma}f_{\sigma}(ia)f_{\beta}(-ia)-\frac{i\alpha^{\prime}}{4(y+ia)}f_{\alpha}(ia)k_{3}^{p}(\gamma_{mp})_{\beta}^{\,\sigma}f_{\sigma}(-ia))]b_{n}(\infty)\rangle.
\end{eqnarray*}

Using residues we arrive at

\begin{eqnarray*}
\mathcal{A} & = & -\frac{\pi\alpha^{\prime}}{18}\xi^{\alpha}\tilde{\xi}^{\beta}a_{2}^{n}a_{3}^{m}\langle[\frac{1}{2}k_{3}^{p}(\gamma_{mp})_{\alpha}^{\,\sigma}f_{\sigma}(ia)f_{\beta}(-ia)+\frac{1}{2}k_{3}^{p}(\gamma_{mp})_{\beta}^{\,\sigma}f_{\alpha}(ia)f_{\sigma}(-ia)]b_{n}(\infty)\rangle
\end{eqnarray*}
and finally our measure give us

\begin{eqnarray*}
\mathcal{A} & = & -\frac{\pi\alpha^{\prime}}{18\times160}\xi^{\alpha}\tilde{\xi}^{\beta}a_{2}^{n}a_{3}^{m}[\frac{1}{2}k_{3}^{p}(\gamma_{mp})_{\alpha}^{\,\sigma}(\gamma_{n})_{\sigma\beta}+\frac{1}{2}k_{3}^{p}(\gamma_{mp})_{\beta}^{\,\sigma}(\gamma_{n})_{\sigma\alpha}].
\end{eqnarray*}

Now using the identity

\begin{eqnarray*}
\gamma_{mp}\gamma_{n}+\gamma_{n}\gamma_{mp}=2\gamma_{nmp}
\end{eqnarray*}
we arrive at

\begin{eqnarray*}
\mathcal{A} & = & -\frac{\pi\alpha^{\prime}}{18\times160}\xi^{\alpha}\tilde{\xi}^{\beta}a_{2}^{n}a_{3}^{m}k_{3}^{p}(\gamma_{nmp})_{\alpha\beta}.
\end{eqnarray*}

From the above we can get the amplitudes for a Kalb-Ramond and two
Photons. Using 

\[
\xi^{\alpha}\widetilde{\xi}^{\beta}=\frac{1}{96}\gamma_{abc}^{\alpha\beta}H^{abc}
\]
we finally get

\begin{eqnarray*}
\mathcal{A} & = & -\frac{\pi}{720\times4\times96}H^{abc}a_{n}^{2}a_{m}^{3}k_{p}^{3}\gamma_{abc}^{\alpha\beta}\gamma_{\alpha\beta}^{nmp}=\frac{i\pi}{720}(\frac{1}{8}a_{a}^{2}F_{bc}^{3}H^{abc})
\end{eqnarray*}
a result which agrees with the previous computation for the Kalb-Ramond and
two photons with pure spinor formalism \cite{Alencar:2008fy}.

\subsubsection{One Gravitino/Dilatino, one Photon and one Photino}

In order to compute this amplitude we must be careful with the construction
of the fixed operator for the gravitino. We need to remember that
this operator comes from the product $\lambda A\bar{\lambda A}$.
Therefore we are going to have two contributions, namely

\begin{eqnarray*}
\frac{1}{18}\left(\xi^{\alpha}f_{\alpha}\tilde{h}^{p}\tilde{b}_{p}+\tilde{\xi}^{\alpha}\tilde{f}_{\alpha}h^{p}b_{p}\right)
\end{eqnarray*}
but because only the zero modes contribute, this is equivalent to consider

\begin{eqnarray*}
\frac{1}{18}(\xi^{\alpha}\tilde{h}^{p}+\tilde{\xi}^{\alpha}h^{p})f_{\alpha}\tilde{b}_{p}=\frac{1}{9}\widetilde{\psi}^{\alpha p}f_{\alpha}\tilde{b}_{p}.
\end{eqnarray*}
 We have considered above the same identification as previously done
in \cite{Alencar:2008fy}. With these considerations the amplitude is given
by

\begin{eqnarray*}
\mathcal{A} & = & \frac{1}{9}\xi^{\alpha}h^{p}\xi_{2}^{\beta}a_{3}^{m}\int dy\langle f_{\alpha}(ia)b_{p}(-ia)f_{\beta}(\infty)[\dot{x}_{m}-ik_{3}^{n}M_{mn}]\rangle.
\end{eqnarray*}

As said before, the first term of the integrated open string gives
a null result. Using now the previous OPEs we get

\begin{eqnarray*}
\mathcal{A} & = & \frac{1}{9}\widetilde{\psi}^{\alpha p}\xi_{2}^{\beta}a_{3}^{m}\int dy\langle\frac{-i\alpha^{\prime}}{4(y-ia)}k_{3}^{n}(\gamma_{mn})_{\alpha}^{\,\sigma}f_{\sigma}(ia))b_{p}(-ia)f_{\beta}(\infty)- \\
 &&i\alpha^{\prime}f_{\alpha}(ia)\frac{k_{3}^{n}{\eta_{np}b_{m}(-ia)-\eta_{mp}b_{n}(-ia)}}{2{y+ia}}f_{\beta}(\infty)\rangle
\end{eqnarray*}
which integrated with residues gives us
\begin{eqnarray*}
\mathcal{A} & = & -\frac{\pi\alpha^{\prime}}{9}\widetilde{\psi}^{\alpha p}\xi_{2}^{\beta}a_{3}^{m}\nonumber \\
&&\langle\frac{1}{2}k_{3}^{n}(\gamma_{mn})_{\alpha}^{\,\sigma}f_{\sigma}(ia)b_{p}(-ia)f_{\beta}(\infty)-f_{\alpha}(ia)k_{3}^{n}(\eta_{np}b_{m}(-ia)-\eta_{mp}b_{n}(-ia))f_{\beta}(\infty)\rangle,
\end{eqnarray*}
and using our measure we arrive at
\begin{eqnarray*}
\mathcal{A} & = & -\frac{\pi\alpha^{\prime}}{9\times160}\widetilde{\psi}^{\alpha p}\xi_{2}^{\beta}a_{3}^{m}[\frac{1}{2}k_{3}^{n}(\gamma_{mn})_{\alpha}^{\,\sigma}\gamma_{p\sigma\beta}-k_{3}^{n}(\eta_{np}\gamma_{m\alpha\beta}-\eta_{mp}\gamma_{n\alpha\beta})].
\end{eqnarray*}

At this point we must remember that we can decompose

\begin{eqnarray*}
\widetilde{\psi}_{p}^{\alpha} & =\psi_{p}^{\alpha} & +\frac{1}{10}\gamma_{p}^{\alpha\delta}\lambda_{\delta}
\end{eqnarray*}
where 

\begin{eqnarray*}
\psi_{p}^{\alpha}=(\widetilde{\psi}_{p}^{\alpha}-\frac{1}{10}\gamma_{p}^{\alpha\delta}\gamma_{\delta\lambda}^{m}\widetilde{\psi}_{m}^{\alpha}), & \lambda_{\delta}=\gamma_{\delta\lambda}^{m}\widetilde{\psi}_{p}^{\alpha}
\end{eqnarray*}
so that $\psi$ is the traceless gravitino and $\lambda$ is the dilatino.
Using this fact, the above amplitude gives us two contributions. 

For the gravitino we have

\begin{eqnarray*}
\mathcal{A} & = & -\frac{\pi\alpha^{\prime}}{9\times160}[\frac{1}{2}\psi^{\alpha p}\xi_{2}^{\beta}a_{3}^{m}k_{3}^{n}(\gamma_{mn})_{\alpha}^{\,\sigma}\gamma_{p\sigma\beta}-\psi^{\alpha p}\xi_{2}^{\beta}a_{3}^{m}k_{3}^{n}(\eta_{np}\gamma_{m\alpha\beta}-\eta_{mp}\gamma_{n\alpha\beta})]
\end{eqnarray*}
and we make use of the identity

\begin{eqnarray*}
\gamma_{mn}\gamma_{p}-\gamma_{p}\gamma_{mn} & = & -2\eta_{np}\gamma_{m}+2\eta_{mp}\gamma_{n}.
\end{eqnarray*}
The second term of the lhs gives a null contribution because of the
gamma traceless property of the gravitino. Therefore we are left with 

\begin{eqnarray*}
\mathcal{A} & = & -\frac{\pi i\alpha^{\prime}}{720}\psi_{m}^{\alpha}\xi_{2}^{\beta}F^{mn}\gamma_{n\alpha\beta}.
\end{eqnarray*}

For the dilatino contribution, we just use the identity

\begin{eqnarray*}
\gamma_{p}\gamma_{mn}\gamma_{p}=6\gamma_{mn}.
\end{eqnarray*}
to get

\begin{eqnarray*}
\mathcal{A} & = & -\frac{\pi i\alpha^{\prime}}{720\times4}F^{mn}\xi_{2}\gamma_{n\alpha\beta}\lambda
\end{eqnarray*}
which agrees with the previous computation for the one gravitino/dilatino,
one photon and one photino with pure spinor formalism \cite{Alencar:2008fy}.

\subsubsection{One Graviton/Dilaton and two Photinos}

In this case we have

\begin{eqnarray*}
\mathcal{A} & = & -\frac{1}{18}h^{p}\widetilde{h}^{m}\xi_{2}^{\alpha}\xi_{3}^{\beta}\int dz\langle[\partial x_{m}-ik_{1}^{n}M_{mn}]b_{p}(-ia)f_{\alpha}(a)f_{\beta}(+\infty)\rangle
\end{eqnarray*}
performing the OPEs we obtain

\begin{eqnarray*}
\mathcal{A} & = & -\frac{1}{18}h^{m}\widetilde{h}^{p}\xi_{2}^{\alpha}\xi_{3}^{\beta}\int dz\langle[\frac{i\alpha^{\prime}}{z-a}k_{m}^{2}f_{\alpha}(a)-ik_{1}^{n}\alpha^{\prime}\frac{(\gamma_{mn})_{\alpha}{}^{\sigma}f_{\sigma}(a)}{4(z-a)}]b_{p}(-ia)f_{\beta}(+\infty)\rangle.
\end{eqnarray*}

Integrating with residues and using our measurement we get

\begin{eqnarray*}
\mathcal{A} & = & \frac{2\pi\alpha^{\prime}}{18\times160}h^{m}\widetilde{h}^{p}\xi_{2}^{\alpha}\xi_{3}^{\beta}[k_{m}^{2}(\gamma_{p})_{\alpha\beta}-\frac{1}{4}k_{1}^{n}(\gamma_{mn})_{\alpha}{}^{\sigma}(\gamma_{p})_{\sigma\beta}].
\end{eqnarray*}

Now by using the equation of motion, the identity

\begin{eqnarray*}
\gamma_{mn}\gamma_{p} & =\gamma_{mnp} & -\eta_{np}\gamma_{m}+\eta_{mp}\gamma_{n}
\end{eqnarray*}
and the fact that just the symmetric part of the polarization contributes,
we get

\begin{eqnarray*}
\mathcal{A} & = & \frac{\pi\alpha^{\prime}}{720\times2}h^{mp}\xi_{2}^{\alpha}\xi_{3}^{\beta}k_{m}^{2}(\gamma_{p})_{\alpha\beta}=-\frac{\pi i\alpha^{\prime}}{720\times2}h^{mp}\xi_{3}\gamma_{p}\partial_{m}\xi_{2}
\end{eqnarray*}
which agrees with the previous computation for the One graviton/dilaton
and two photinos with pure spinor formalism \cite{Alencar:2008fy}.

\subsection{One Graviton/Dilaton and two Photons}

This is the last case to be considered. The amplitude is given by

\begin{eqnarray*}
\mathcal{A} & = & h^{p}\widetilde{h}^{q}a_{2}^{r}a_{3}^{m}\int dy\langle V_{p}^{B}(ia)\overline{V}_{q}^{B}(-ia)V_{r}^{B}(\infty)U_{m}^{B}(y)\rangle.
\end{eqnarray*}

As argued before, this will give rise to three terms. The first one
is given by

\begin{eqnarray*}
\mathcal{A}_{1}  & =-\frac{1}{16}h^{p}\widetilde{h}^{q}a_{2}^{r}a_{3}^{m}\int dy\langle\frac{i}{2}k_{1}^{s}(\gamma_{ps})_{\alpha}{}^{\sigma}V_{\sigma}^{F}(ia)\gamma_{q}^{\alpha\beta}V_{\beta}^{F}(-ia)V_{r}^{B}(\infty)U_{m}^{B}(y)\rangle,
\end{eqnarray*}
and using our previous result for the Kalb-Ramond and two photons
we get
\begin{eqnarray*}
\mathcal{A}_{1}=-\frac{\pi\alpha^{\prime}}{16\times18\times160}h^{p}\widetilde{h}^{q}a_{2}^{r}a_{3}^{m}\frac{i}{2}k_{1}^{s}(\gamma_{ps})_{\alpha}{}^{\sigma}\gamma_{q}^{\alpha\beta}k_{3}^{t}(\gamma_{mtr})_{\sigma\beta}.
\end{eqnarray*}

Using now the identity
\begin{eqnarray*}
Tr(\gamma^{q}\gamma^{ps}\gamma_{mtr})=-96\delta_{mrt}^{pqs}
\end{eqnarray*}
and the fact that the metric is symmetric we see that the above gives
a null result. 

The next term is given by

\begin{eqnarray*}
\mathcal{A}_{2} & = & -\frac{1}{16}\widetilde{h}^{q}a_{2}^{r}a_{3}^{m}\int dy\langle V_{p}^{B}(ia)\gamma_{q}^{\alpha\beta}V_{\beta}^{F}(-ia)\{q_{\alpha},V_{r}^{B}(\infty)\}U_{m}^{B}(y)\rangle
\end{eqnarray*}

\begin{eqnarray*}
=-\frac{1}{16}h^{p}\widetilde{h}^{q}a_{2}^{r}a_{3}^{m}\int dy\langle V_{p}^{B}(ia)\gamma_{q}^{\alpha\beta}V_{\beta}^{F}(-ia)\frac{i}{2}k_{2}^{s}(\gamma_{rs})_{\alpha}{}^{\sigma}V_{\sigma}^{F}(\infty)U_{m}^{B}(y)\rangle
\end{eqnarray*}
and using our previous result we get

\begin{eqnarray*}
\mathcal{A}_{2} & = & -\frac{\pi\alpha^{\prime}}{16\times18\times160}h^{p}\widetilde{h}_{q}a_{2}^{r}a_{m}^{3}(\gamma^{q})^{\alpha\beta}\frac{i}{2}k_{2}^{s}(\gamma_{rs})_{\alpha}{}^{\sigma}[\frac{1}{2}ik_{n}^{3}(\gamma^{mn})_{\beta}^{\,\delta}\gamma_{p\delta\sigma}-ik_{n}^{3}(\delta_{p}^{n}\gamma_{m\beta\sigma}-\delta_{p}^{m}\gamma_{n\beta\sigma})].
\end{eqnarray*}
Using the identities

\begin{eqnarray*}
\gamma^{q}\gamma_{rs}\gamma_{p}\gamma^{mn} & =-32\delta_{s}^{q}\delta_{pr}^{mn}+32\delta_{r}^{q}\delta_{ps}^{mn}-32\delta_{s}^{p}\delta_{qr}^{mn}+32\delta_{r}^{p}\delta_{qs}^{mn}-32\delta_{q}^{p}\delta_{rs}^{mn}
\end{eqnarray*}
and

\begin{eqnarray*}
\gamma^{q}\gamma_{rs}\gamma^{m} & =32\delta_{rs}^{mq}
\end{eqnarray*}
and using the fact that $h_{p}^{p}=4\Phi$ we finally get

\begin{eqnarray*}
\mathcal{A}_{2} & = & \frac{\pi i\alpha^{\prime}}{720}[-\frac{1}{4}h_{q}^{p}F_{2}^{rq}F_{rp}^{3}-\frac{1}{8}\Phi F_{2}^{rs}F_{rs}^{3}].
\end{eqnarray*}

The last term is given by

\begin{eqnarray*}
\mathcal{A}_{3} & = & -\frac{1}{16}h^{p}\widetilde{h}^{q}a_{2}^{r}a_{3}^{m}\int dy\langle V_{p}^{B}(ia)\gamma_{q}^{\alpha\beta}V_{\beta}^{F}(-ia)V_{r}^{B}(\infty)\{q_{\alpha},U_{m}^{B}(y)\}\rangle
\end{eqnarray*}

\begin{eqnarray*}
= &  & -\frac{1}{16}h^{p}\widetilde{h}^{q}a_{2}^{r}a_{3}^{m}\int dy\langle V_{p}^{B}(ia)\gamma_{q}^{\alpha\beta}V_{\beta}^{F}(-ia)V_{r}^{B}(\infty)\frac{i}{2}k_{3}^{n}(\gamma_{mn})_{\alpha}{}^{\sigma}U_{\sigma}^{F}\rangle
\end{eqnarray*}
and using the cyclic symmetry we get 

\begin{eqnarray*}
\mathcal{A}_{3} & = & -\frac{1}{16}h^{p}\widetilde{h}^{q}a_{2}^{r}a_{3}^{m}\frac{i}{2}\int dy\langle V_{p}^{B}(ia)\gamma_{q}^{\alpha\beta}V_{\beta}^{F}(-ia)U_{r}^{B}(y)k_{3}^{s}(\gamma_{ms})_{\alpha}{}^{\sigma}V_{\sigma}^{F}(\infty)\rangle.
\end{eqnarray*}

This is identical to our last term if we exchange $k_{1}\leftrightarrow k_{2}$.
Therefore we get 

\begin{eqnarray*}
\mathcal{A}_{3} & = & \frac{\pi i\alpha^{\prime}}{720}[-\frac{1}{4}h_{q}^{p}F_{3}^{rq}F_{rp}^{2}-\frac{1}{8}\Phi F_{3}^{rs}F_{rs}^{2}],
\end{eqnarray*}
and adding the results we obtain

\begin{eqnarray*}
\mathcal{A}_{3} & = & \frac{\pi i\alpha^{\prime}}{720}[-\frac{1}{2}h_{q}^{p}F_{3}^{rq}F_{rp}^{2}-\frac{1}{4}\Phi F_{3}^{rs}F_{rs}^{2}]
\end{eqnarray*}
which finally agrees with the previous computation for the one graviton/dilaton
and two photons with pure spinor formalism \cite{Alencar:2008fy}.

\section{Conclusions}

In this manuscript we have made a proof for the equivalence between the RNS
and Pure Spinor formalism for a class of Type I tree level amplitudes.
For this we have constructed a proof for the cyclic symmetry of
superstring Type I amplitudes using a BRST approach. This symmetry
was used previously in Ref.\cite{Alencar:2008fy}.
That was stablished for amplitudes involving up to four fermions. With
this at hand, we have all the terms of the Type I effective action
obtained using the RNS formalism which involves these kind of amplitudes.
The amplitude involving one closed and two open strings was computed
previously \cite{Alencar:2008fy} with the Pure Spinor formalism. This result was shown
to be correct by comparing with the effective action for Type I Supergravity.
With the present result explicit computations are not needed anymore.
This result also give us a result of particular importance. Differently
from amplitudes involving only closed or open strings, the Type I
superstring has a gauge anomaly that is manifest already at tree level.
Diagrams in which a two form is exchanged between two gauge fields
on one side and four on the other side have to be considered for this
cancellation. This amplitude involves two fermions and six bosons
and so is obtained from our present result. Therefore we get as a
byproduct of this manuscript the computation of this amplitude. 

Recently, some important results regarding amplitude computations have
been obtained. First we can mention a relation between
disk amplitudes involving $N_{o}$ open and $N_{c}$ closed strings
and disk amplitudes with only $N_{o}+2N_{c}$ open strings\cite{Stieberger:2009hq}. With this, and our results
we can obtain many explicit amplitudes from the already known for
Pure Spinor. Between these we can cite the six point amplitude computed
recently \cite{Mafra:2010gj}, that could give us all the terms of the effective
action involving one closed and four open or two closed and two
open. Results of a more far reaching importance is the computation of any
number of open strings at tree level \cite{Mafra:2010jq}. With this we also
gain as a byproduct the computation of any amplitude involving any
number of open and closed string that contain up to four fermions.
Posteriorly the result above was used to compute Supergravity and
Super-Yang-Mills amplitudes. As a perspective of the present work
the authors expect to reach similar results involving Type I Supergravity
amplitudes.

\section*{Acknowledgment}

The authors would like to thank Brenno Carlini Vallilo, Nathan Berkovits and C. R. Mafra for useful discussions. We would also like to thank: The Goethe-Institut Berlin for the hospitality. 
We acknowledge Linda Fuchs for her help with the manuscript, and the financial
support provided by Funda\c c\~ao Cearense de Apoio ao Desenvolvimento Cient\'\i fico e Tecnol\'ogico (FUNCAP), the Conselho Nacional de Desenvolvimento Cient\'\i fico e Tecnol\'ogico (CNPq) and FUNCAP/CNPq/PRONEX.

\end{document}